\documentclass{article}

\PassOptionsToPackage{numbers}{natbib}
\usepackage[preprint]{neurips_2021}

\usepackage[utf8]{inputenc} 
\usepackage[T1]{fontenc}    
\usepackage{booktabs}       
\usepackage{amsfonts}       
\usepackage{nicefrac}       
\usepackage{microtype}      
\usepackage{amsmath}
\usepackage{graphicx}
\usepackage{tipa}
\usepackage{multirow}
\usepackage{longtable}

\usepackage{subcaption}
\usepackage{color}
\usepackage[table]{xcolor}
\usepackage{threeparttable}
\usepackage{diagbox}
\usepackage{enumitem}
\usepackage{tablefootnote}
\usepackage{hyperref}
\usepackage{url}            
\usepackage{xurl}
\usepackage{breakurl}

\usepackage[edges]{forest}
\usepackage{tikz}
\usepackage{tikz-qtree}
\usetikzlibrary{fadings}
\usetikzlibrary{mindmap,trees}
\usetikzlibrary{arrows,automata,shapes,positioning,shadows,trees}
\usetikzlibrary{shapes,snakes,shadows}
\usetikzlibrary{shapes.arrows}
\usetikzlibrary{calc,shapes, positioning}
\usetikzlibrary{decorations.text}
\usetikzlibrary{matrix,chains,positioning,decorations.pathreplacing,arrows}
\usetikzlibrary{bayesnet}
\usetikzlibrary{shadows.blur}
\usetikzlibrary{shapes.geometric}

\tikzstyle{edge}=[-latex',draw=black!90,shorten <=1pt,shorten >=1pt]
\tikzstyle{redge}=[latex'-,draw=black!90,shorten <=1pt,shorten >=1pt]
\tikzstyle{dedge}=[latex'-latex',draw=black!90,shorten <=1pt,shorten >=1pt]
\tikzstyle{block}=[draw, text width=5em,align=center,shape=rectangle, rounded corners, , align=center]
\tikzstyle{nobox}=[align=center]

\definecolor{emb}{RGB}{209,228,252}
\definecolor{hidden-blue}{RGB}{194,232,247}
\definecolor{hidden-orange}{RGB}{243,202,120}
\definecolor{hidden-yellow}{RGB}{242,244,193}
\definecolor{output-purple}{RGB}{219,203,231}
\definecolor{output-green}{RGB}{204,231,207}
\definecolor{hiddendraw}{RGB}{205, 44, 36}

\tikzstyle{mybox}=[
    rectangle,
    draw=hiddendraw,
    rounded corners,
    text opacity=1,
    minimum height=1.5em,
    minimum width=5em,
    inner sep=2pt,
    align=center,
    fill opacity=.2,
    ]

\tikzstyle{emb-purple}=[
    rectangle,
    draw=output-purple!50!purple,
    fill=output-purple,
    text opacity=1,
    minimum height=1.5em,
    minimum width=1.5em,
    inner sep=0pt,
    align=center,
    fill opacity=.9,
    ]

\tikzstyle{emb-blue}=[
    rectangle,
    draw=emb!50!blue,
    fill=emb,
    text opacity=1,
    minimum height=1.5em,
    minimum width=1.5em,
    inner sep=0pt,
    align=center,
    fill opacity=.9,
]

\definecolor{colone}{RGB}{178, 34, 34}
\definecolor{coltwo}{RGB}{106, 90, 205}
\definecolor{colthree}{RGB}{255, 250, 205}
\definecolor{colfour}{RGB}{0, 139, 69}
\definecolor{colfive}{RGB}{245,238,197}
\definecolor{colsix}{RGB}{243,235,179}
\definecolor{colseven}{RGB}{241,231,163}
\title{A Survey on Audio Synthesis and Audio-Visual Multimodal Processing}

\author{
Zhaofeng Shi\\
\texttt{2017020912015@std.uestc.edu.cn} \\
University of Electronic Science and Technology of China \\
}

\begin{document}
\maketitle

\begin{abstract}

With the development of deep learning and artificial intelligence, audio synthesis has a pivotal role in the area of machine learning and shows strong applicability in the industry. Meanwhile, significant efforts have been dedicated by researchers to handle multimodal tasks at present such as audio-visual multimodal processing. In this paper, we conduct a survey on audio synthesis and audio-visual multimodal processing, which helps understand current research and future trends. This review focuses on text to speech(TTS), music generation and some tasks that combine visual and acoustic information. The corresponding technical methods are comprehensively classified and introduced, and their future development trends are prospected. This survey can provide some guidance for researchers who are interested in the areas like audio synthesis and audio-visual multimodal processing.

\end{abstract}

\section{introduction}
\label{inrtoduction}
Audio synthesis, which aims to synthesis various form of natural and intelligible sound such as speech, music, has a wide range of application scenario in human society and industry. Initially, researchers took advantages of pure signal processing methods to find some convenient representations for audio, which can be easily modelled and transform to temporal audio. For example, short-time Fourier Transform(STFT) is an efficient way to convert audio into frequency domain and Griffin-Lim~\cite{griffin1984signal} is a kind of pure signal processing algorithm that is able to decode STFT sequence to temporal waveform. Methods similar to Griffin-Lim are WORLD~\cite{morise2016world}, etc. In recent years, with the rapid development of deep learning technology, researchers have begun to build deep neural networks for audio synthesis and other multimodal tasks in order to simplify the pipeline and improve the performance of the model. Numerous neural network models have emerged so far for the tasks such as text to speech(TTS) and music generation. There are a lot of models for TTS that have been reported, such as Parallel WaveGAN~\cite{yamamoto2020parallel}, MelGAN~\cite{kumar2019melgan}, FastSpeech2/2s~\cite{ren2020fastspeech}, EATs~\cite{donahue2020end}, VITS~\cite{kim2021conditional}. Simultaneously, there are many models for music generation like song from PI~\cite{chu2016song}, C-RNN-GAN~\cite{mogren2016c}, MidiNet~\cite{yang2017midinet}, MuseGAN~\cite{dong2018musegan} and Jukebox~\cite{dhariwal2020jukebox}. These models bring great convenience to human production and life, and they provide key reference for future research. 

Vision is a physiological word. Humans and animals visually perceive the size, brightness, color, etc. of external objects, and obtain information that is essential for survival. Vision is the most important sense for human beings. Over these years, deep learning has been widely explored in various image processing and computer vision tasks such as image dehazing/deraining, objective detection and image segmentation, which contribute to the development of social productivity. Image dehazing/deraining means given a blurred image with haze/rain, algorithms are used to remove the haze/rain in the image to make it clear. \cite{wei2020single,li2020region,wu2020subjective,li2020haze,wu2020unified,luo2021single,wu2019beyond,wei2021non,li2021single} proposed neural network-based models for image dehazing/deraining respectively. Objective detection means finding out all the targets of interest in the image and determining their locations and classes. In~\cite{ding2020human,qiu2020hierarchical,qiu2020offset,wang2020multi,li2020codan,li2019simultaneously,li2019headnet,qiu2019a2rmnet,chen2021bal,chen2020high,li2018incremental}, several neural network-based models with high performance for objective detection can be found. Image segmentation means dividing an image into a number of specific regions with unique properties and presenting objects of interest. The work on image segmentation includes~\cite{yang2020new,guo2020deep,shi2020query,yang2020learning,yang2020mono,meng2019new,yang2019new,xu2019bounding,shang2019instance,luo2018weakly,shi2018key}.

Modality means the form or source of every kind of information such as visual, auditory and tactile. Information from different modalities is often closely related in our life. For example, when talking, we need to combine the other’s facial expression and speech to determine his emotions. With the development of deep learning, multimodal machine learning has become a research hotspot. In this review, we focus on audio-visual multimodal processing, which has a large range of applications. Many well-performing models for audio-visual multimodal processing has been proposed in the past few years. For instance, Vid2Speech~\cite{ephrat2017vid2speech} reported by Ephrat et al. is able to handle lipreading task. Acapella~\cite{montesinos2021cappella} is a model for singing voice separation. VisualSpeech~\cite{gao2021visualvoice} is a versatile model, which achieves state of the art performance in speech separation and speech enhancement tasks. More models for audio-visual multimodal processing will be comprehensively summarized in section~\ref{Audio-visual multimodal processing}.

In this review, we mainly review the research work on three aspects, TTS, music generation and audio-visual multimodal processing based on deep learning technology. In section~\ref{Text to speech}, section~\ref{Music generation} and section~\ref{Audio-visual multimodal processing}, we introduce the background and technical methods corresponding to the three tasks respectively. In section~\ref{Datasets}, we present several of the most commonly used datasets in studies of different areas respectively. Finally, in section~\ref{Conclusion}, we present conclusions based on the description and discussion in the above. The contributions of this review include:
\begin{itemize}[leftmargin=*]
    \item We introduce and summarize the background and technical methods of TTS, music generation and multimodal audio-visual processing tasks.
    \item This review provides some guidance for researchers who are interested in this direction.
\end{itemize}

\tikzstyle{leaf}=[mybox,minimum height=1.2em,
fill=hidden-orange!50, text width=5em,  text=black,align=left,font=\footnotesize,
inner xsep=4pt,
inner ysep=1pt,
]

\begin{figure*}[thp]
 \centering
\begin{forest}
  forked edges,
  for tree={
  grow=east,
  reversed=true,  
  anchor=base west,
  parent anchor=east,
  child anchor=west,
  base=left,
  font=\normalsize,
  rectangle,
  draw=hiddendraw,
  rounded corners,
  align=left,
  minimum width=2.5em,
  inner xsep=4pt,
  inner ysep=0pt,
  },
    [Survey
        [Sec.~\ref{Text to speech}: Text to speech
            [Sec.~\ref{Two-stage methods}: Two-stage methods
                [Sec.~\ref{Acoustic models}: Acoustic models
                ]
                [Sec.~\ref{Vocoders}: Vocoders
                ]
            ]
            [Sec.~\ref{End-to-End methods}: End-to-End methods
            ]
        ]
        [sec.~\ref{Music generation}: Music generation
            [Sec.~\ref{Symbolic music generation}: Symbolic music generation
            ]
            [Sec.~\ref{Audio music generation}: Audio music generation
            ]
        ]
        [sec.~\ref{Audio-visual multimodal processing}: Audio-visual multimodal processing
            [sec.~\ref{Lipreading}: Lipreading
            ]
            [sec.~\ref{Audio-visual speech separation}: Audio-visual speech separation
            ]
            [sec.~\ref{Talking face generation}: Talking face generation
            ]
            [sec.~\ref{Sound generation from video}: Sound generation from video
            ]
        ]
        [sec.~\ref{Datasets}: Datasets
            [sec.~\ref{Datasets for TTS}: Datasets for TTS
            ]
            [sec.~\ref{Datasets for music generation}: Datasets for music generation
            ]
            [sec.~\ref{Datasets for audio-visual multimodal processing}: Datasets for audio-visual multimodal processing
            ]
        ]
    ]
\end{forest}
\caption{Organization of this paper.}
\label{org_survey_paper}
\end{figure*}
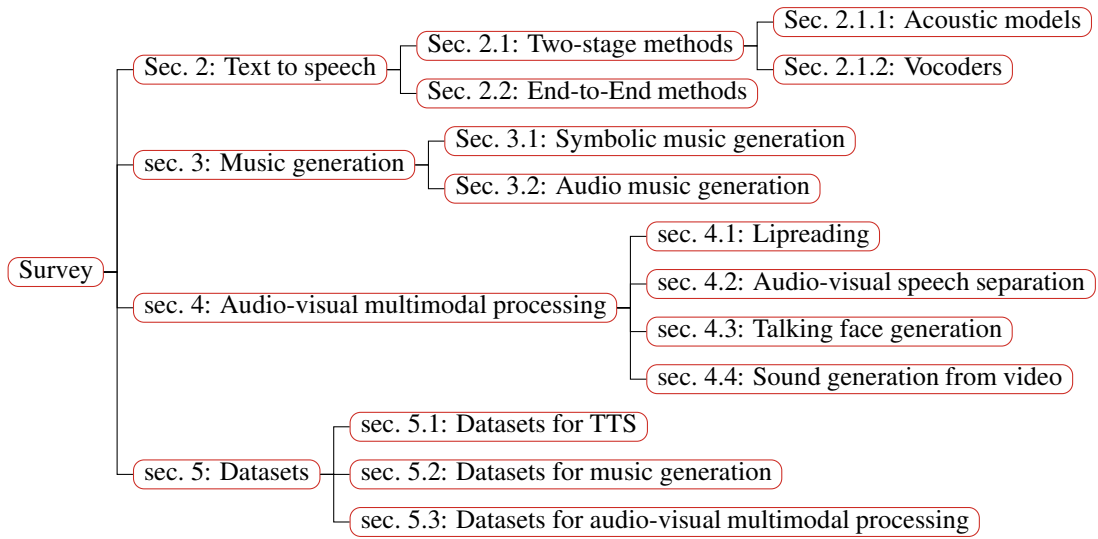

\section{Text to Speech}
\label{Text to speech}
Text to speech, also known as speech synthesis, which aims to generate natural and intelligible speech waveform under the condition of the input text~\cite{tan2021survey}. With the continuous development of artificial intelligence technology, TTS has gradually become a hot research topic and it has been widely adopted in the field of audiobook services, automotive navigation services, automated answering services, etc. However, the requirements of the naturalness and accuracy of the speech generated by the TTS system are also increasing as well. A well-designed neural network is needed in order to meet these requirements.

Modelling the raw audio waveform is a challenging problem because the temporal resolution of the audio data is extremely high. The sampling rate of audio is at least 16000Hz. And another challenge is that there are long-term and short-term dependencies in the audio waveform, which are difficult to model. In addition to this, there are other difficulties for the TTS tasks such as the alignment problem and the one-to-many problem. The alignment problem means that the output speech audio content should be temporal aligned with the input text or phoneme sequences. At present, the mainstream solution to this problem is to adopt Monotonic Alignment Search(MAS)~\cite{kim2020glow}. As for the one-to-many problem, it means that the text input can be spoken in multiple ways with different variations such as duration, pitch, and there are many kinds of neural network structures have been designed to deal with this problem.

The frameworks of TTS tasks are mainly based on two-stage methods and end-to-end methods respectively. Most previous work construct two-stage pipelines apart from text preprocessing such as text normalization and conversion from text to phoneme. The first stage is to produce intermediate representations such as linguistic features~\cite{oord2016wavenet} or acoustic features such as mel-spectrograms~\cite{shen2018natural} under the condition of the input sequence consist of phonemes, characters, etc. The model of the second stage is also known as vocoder, which generates raw audio waveforms conditioned on the input intermediate representations. Numerous models of each part of the two-stage pipelines have been developed independently. Taking advantage of the two-stage method, we can get high-fidelity audio. Nonetheless, two-stage pipelines remain problematic because the models need fine-tuning and the errors of the previous module may affect the next module causing error accumulation. To solve this problem, more and more end-to-end models have been developed by researchers. End-to-end models output raw audio waveforms directly without any intermediate representations. Because it can parallel generate audio in a short time and alleviate the problem of error accumulation without spending a lot of time tuning parameters, the end-to-end method has become the current research direction in TTS field. The details of these two methods will be introduced below.

\subsection{Two-stage methods}
\label{Two-stage methods}
As mentioned above, the first stage of the two-stage method is to convert the input text sequence into the intermediate representation and the second stage generates raw audio waveform. In the first stage, a linguistic model or an acoustic model is deployed to generate linguistic features or acoustic features as intermediate features. The model in the second stage, which is also called vocoder, converts the intermediate representations into the raw audio waveform. To generate high-quality speech audio, these two models must operate well together. The models corresponding to two-stage methods will be introduced and summarized below.

\subsubsection{Acoustic models}
\label{Acoustic models}
Nowadays, acoustic features are usually used as the intermediate features in TTS tasks. As a result, we focus on the research work on acoustic models in this section. Acoustic features are generated by acoustic models from linguistic features or directly from characters or phonemes~\cite{tan2021survey}. The aim of acoustic models is to generate acoustic features that are further converted into raw audio waveform using vocoders in two-stage methods of TTS tasks. The acoustic models that have been reported in recent years will be comprehensively summarized below.

Tacotron and Tacotron 2~\cite{wang2017tacotron,shen2018natural} are autoregressive neural network-based models that take raw text characters as input and output mel-spectrograms. Other autoregressive models similar to them, which take mel-spectrograms as output, including DeepVoice 3, Transformer TTS, Flowtron, MelNet~\cite{ping2017deep,li2019neural,valle2020flowtron,vasquez2019melnet}, etc. It is worth mentioning that MelNet is an autoregressive model that takes both time and frequency into consideration. After the knowledge distillation~\cite{hinton2015distilling} was proposed by Geoffrey et al., more and more researchers have tried to take advantage of knowledge distillation to train acoustic models like ParaNet~\cite{peng2020non} and Fastspeech~\cite{ren2019fastspeech}. They are both non-autoregressive models. In addition, Flow-TTS~\cite{miao2020flow}, Glow-TTS~\cite{kim2020glow}, etc. are flow-based non-autoregressive models and Fastspeech 2~\cite{ren2020fastspeech} is a non-autoregressive model based on Transformer~\cite{vaswani2017attention}.

\subsubsection{Vocoders}
\label{Vocoders}
Vocoder is the model that converts the intermediate representations produced in the first stage into the raw audio waveform. In recent years, many neural network-based vocoders have been designed to achieve high-fidelity speech audio synthesis. We will summarize the research work related to vocoders below, and classify these models into two categories: autoregressive models and non-autoregressive models.

\paragraph{Autoregressive models}
WaveNet~\cite{oord2016wavenet} is the first neural network-based vocoder, which takes linguistic features as input and generated raw audio waveform autoregressively. WaveNet is fully convolutional. By using dilated causal convolutions, WaveNet can achieve a large receptive field, which plays an important role in addressing the issue of modelling the high temporal resolution audio waveform. SampleRNN~\cite{mehri2016samplernn} is another autoregressive model based on RNN. It is an unconditional end-to-end neural audio generative model, which uses a hierarchy of RNN module, each operating at a different time resolution to deal with the challenge of modelling the raw audio waveform at different temporal resolution. DeepVoice~\cite{arik2017deep} and LPCNet~\cite{valin2019lpcnet} are also autoregressive models, like the two models mentioned above. However, for autoregressive models, each output sample point is conditioned on the previously generated sample points. As a result, inference with these models is time-consuming because audio sample points must be generated sequentially. In response to these problems, researchers designed non-autoregressive models that can generate audio in parallel.

\paragraph{Non-autogressive models}
Non-autoregressive models can generate raw audio samples in parallel to quickly generate speech audio to satisfy real-time requirements. There are many technical methods that have been proven to be applicable to non-autoregressive models. WaveGlow, FloWaveNet and WaveFlow~\cite{prenger2019waveglow,kim2018flowavenet,ping2020waveflow} are flow-based non-autoregressive models. Parallel WaveNet~\cite{oord2018parallel} is also a flow-based model, which combines knowledge distillation~\cite{hinton2015distilling} to allow the inverse autoregressive flow(IAF)~\cite{kingma2016improved} network as a student network and the pretrained WaveNet as a teacher network. ClariNet~\cite{ping2018clarinet} is also a vocoder that employs knowledge distillation~\cite{hinton2015distilling}. However, the training process with distillation-based methods remain problems. It requires not only a robust teacher model, but also a methodology to optimize the complicated distillation process. Much recent progress in Generative Adversarial Networks(GANs)~\cite{2014Generative} has been expedited and WaveGAN~\cite{donahue2018adversarial} is the first GAN-based vocoder, which modifies the structure of DCGAN~\cite{radford2015unsupervised} to realize audio synthesis. In addition, Parallel WaveGAN~\cite{yamamoto2020parallel} is a GAN-based model using multi resolution short time Fourier Transform(STFT) to capture the time-frequency distribution of the natural speech audio. MelGAN~\cite{kumar2019melgan} is another non-autoregressive feed-forward vocoder based on GAN, which inverts the input mel-spectrogram into raw speech audio. Other GAN-based vocoders include GAN-TTS~\cite{binkowski2019high}, HiFi-GAN~\cite{kong2020hifi}, etc.

\tikzstyle{leaf}=[mybox,minimum height=1.2em,
fill=hidden-orange!50, text width=5em,  text=black,align=left,font=\footnotesize,
inner xsep=4pt,
inner ysep=1pt,
]

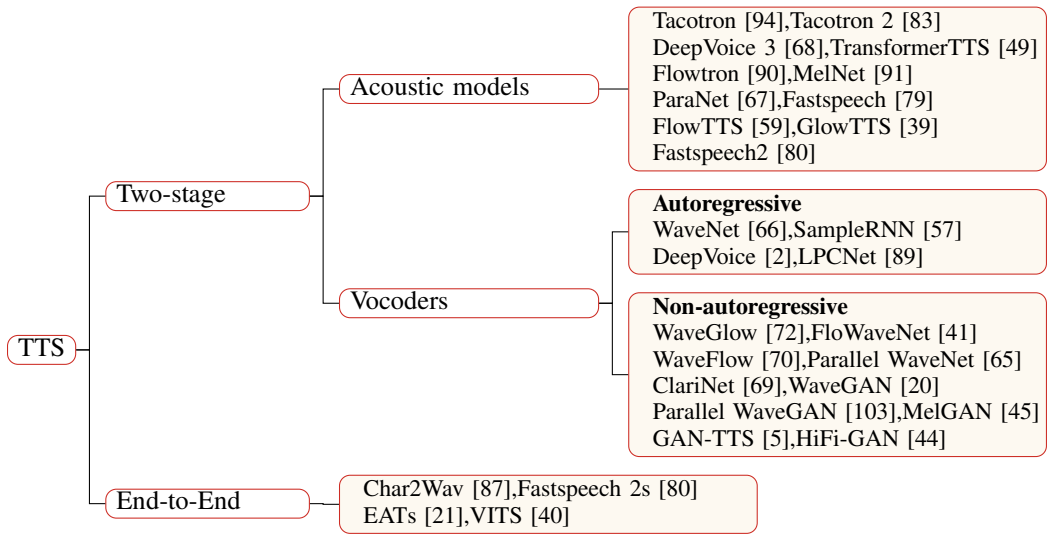
\begin{figure*}[thp]
 \centering

\begin{forest}
  forked edges,
  for tree={
  grow=east,
  reversed=true,  
  anchor=base west,
  parent anchor=east,
  child anchor=west,
  base=left,
  font=\normalsize,
  rectangle,
  draw=hiddendraw,
  rounded corners,
  align=left,
  minimum width=2.5em,
  inner xsep=4pt,
  inner ysep=0pt,
  },
  where level=1{text width=6.9em,font=\normalsize}{},
  where level=2{text width=9em,font=\normalsize}{},
  where level=3{font=\footnotesize,yshift=0.25pt}{},
    [TTS
        [Two-stage
            [Acoustic models
                [Tacotron~\cite{wang2017tacotron}{,}Tacotron 2~\cite{shen2018natural}\\
                DeepVoice 3~\cite{ping2017deep}{,}TransformerTTS~\cite{li2019neural}\\
                Flowtron~\cite{valle2020flowtron}{,}MelNet~\cite{vasquez2019melnet}\\
                ParaNet~\cite{peng2020non}{,}Fastspeech~\cite{ren2019fastspeech}\\
                FlowTTS~\cite{miao2020flow}{,}GlowTTS~\cite{kim2020glow}\\
                Fastspeech2~\cite{ren2020fastspeech}
                ,leaf,text width=15em
                ]
            ]
            [Vocoders
                [\textbf{Autoregressive}\\
                WaveNet~\cite{oord2016wavenet}{,}SampleRNN~\cite{mehri2016samplernn}\\
                DeepVoice~\cite{arik2017deep}{,}LPCNet~\cite{valin2019lpcnet}
                ,leaf,text width=15em
                ]
                [\textbf{Non-autoregressive}\\
                WaveGlow~\cite{prenger2019waveglow}{,}FloWaveNet~\cite{kim2018flowavenet}\\
                WaveFlow~\cite{ping2020waveflow}{,}Parallel WaveNet~\cite{oord2018parallel}\\
                ClariNet~\cite{ping2018clarinet}{,}WaveGAN~\cite{donahue2018adversarial}\\
                Parallel WaveGAN~\cite{yamamoto2020parallel}{,}MelGAN~\cite{kumar2019melgan}\\
                GAN-TTS~\cite{binkowski2019high}{,}HiFi-GAN~\cite{kong2020hifi}
                ,leaf,text width=15em
                ]
            ]
        ]
        [End-to-End
            [Char2Wav~\cite{sotelo2017char2wav}{,}Fastspeech 2s~\cite{ren2020fastspeech}\\
            EATs~\cite{donahue2020end}{,}VITS~\cite{kim2021conditional}
            ,leaf,text width=15em
            ]
        ]
    ]
\end{forest}
\caption{A taxonomy of TTS.}
\label{main_taxonomy_of_tts}
\end{figure*}

\subsection{End-to-End methods}
\label{End-to-End methods}
End-to-end TTS models can directly convert the input text or phoneme sequence into speech audio waveform without any intermediate representations and this kind of framework admits many straightforward extensions. Fully end-to-end models can not only reduce the time cost of training and reference, but also avoid error propagation in cascade models. Moreover, it also requires less human annotations in the datasets. However, these advantages come with challenges. The first thing is the huge difference between modalities of text and waveform. For instance, the temporal resolution of text and waveform is very different: there are 5 words per second on average during speech, the length of corresponding phoneme sequence is just around 20. But the sample rate of speech audio is usually as high as 16000Hz or even higher. Extremely large difference in temporal resolution makes it difficult to model text-audio correlations. And the requirement of memory is high because of high temporal resolution of audio. Besides, the lack of aligned intermediate representations makes the alignment between text and audio more difficult. In response to these challenges, researchers have designed a variety of end-to-end TTS networks, which achieve excellent performance.

Char2Wav~\cite{sotelo2017char2wav} is the first end-to-end model based on RNN and various end-to-end solutions emerged after it was reported. Fastspeech 2s~\cite{ren2020fastspeech} is a GAN-based non-autoregressive model that uses the feed-forward Transformer~\cite{vaswani2017attention} block and 1D-convolution as the basic structure for the encoder and decoder. In addition, the variance adaptor was designed to deal with the one-to-many problem and facilitate the alignment of text and speech. EATs~\cite{donahue2020end} is a GAN-based non-autoregressive end-to-end TTS model like Fastspeech 2s. A well-designed aligner is the key part of the whole framework to obtain the aligned representation from the input phoneme sequence and decode it into raw audio waveform. Dynamic time warping is applied to solve the one-to-many problem. VITS~\cite{kim2021conditional} is a TTS model based on flow and Variational Autoencoder(VAE)~\cite{kingma2013auto}, which adopts Monotonic Alignment Search(MAS)~\cite{kim2020glow} and stochastic duration predictor to address the alignment and one-to-many problems respectively and achieves state-of-the-art performance.

Although there have been many two-stage TTS models with excellent performance, end-to-end models has become a hot research topic in recent years because of their efficiency and accuracy. Hence, high-efficiency end-to-end models will become a trend of future research in TTS area.

\section{Music generation}
\label{Music generation}
In recent years, researchers have begun to try to take advantage of artificial intelligence generating realistic and aesthetic pieces such as images, text and videos. Music is an art that reflects the emotions of human beings in real life. The use of computer technology for algorithmic composition dates back to the last century. With the development of deep learning technology, generating music by deep neural networks 
has become a trending research topic. Using deep learning technology to generate music is meaningful. In life, people without any music theory knowledge can also use the technology to generate music to relax themselves. In the commercial field, large-scale tailer-made music generation can be realized with low time and economic costs through the technology. In addition, neural network-based music generation technology has broad prospects in education, medical and other fields. Hence, it is a promising research topic.

Different from image generation and text generation, music generation has its unique challenges. Firstly, music is an art of time~\cite{dong2018musegan}. Music is hierarchical, and there are usually long-term and short-term dependencies within a piece of music. Therefore, the hierarchical structure and temporal dependencies in music must be well modeled. Secondly, Music usually consists of multiple tracks and instruments. It is necessary to organically compose different tracks and instruments to generate natural, pleasant music. Thirdly, there are chords and melodies in music, which consist of a series of notes arranged in a specific order. The formation of chords and melodies plays a pivotal role in music generation. In summary, the three points mentioned above should be considered carefully when designing the structure of neural networks for music generation.

Based on the current research status of neural network-based music generation, music generation approaches can be divided into two classes: symbolic music generation and audio music generation. Symbolic music generation means that music is symbolically generated in the form of a sequence of notes, which specifies information such as pitch, duration and instrument. These notes further compose melodies and chords and the generated music is stored in midi format finally. Taking advantage of this approach, music can be modeled in the low-dimensional space, which can significantly reduce the difficulty of music generation task. Moreover, since music is composed of a series of musical events, no harsh noise is introduced. As for audio music generation, it means that music is generated at the audio sample point level. The sample rate of music audio waveform is usually 44100Hz. Therefore, the key bottleneck of audio music generation is that modeling long-term and short-term dependencies at such a high temporal resolution. This approach has some advantages. Firstly, directly generating raw audio waveform of music can avoid a large number of complicated human annotations of datasets. Secondly, audio music generation can generate unprecedented sounds and even voice from singer instead of being limited to several specific musical instruments. We will introduce research work related to these two approaches below, respectively.

\subsection{Symbolic music generation}
\label{Symbolic music generation}
Symbolic music generation means that the model generates a series of musical events such as note, instrument, pitch and duration and composes them into music in midi format. Nowadays, some neural network-based symbolic music generative models have been proposed, which can be divided into two classes according to the network structure: RNN-based models and CNN-based models. We will present several typical models in each of these two classes, respectively.

\paragraph{RNN-based models}
Many excellent generative models of symbolic-domain music have been proposed up to now. DeepBach~\cite{hadjeres2017deepbach} is an RNN-based model, which is capable of generating highly natural Bach-style chorales. Song from PI~\cite{chu2016song} is a four-layer hierarchical RNN model to generate symbolic pop music, with different layers corresponding to Drum, Chord, Press and Key, respectively. C-RNN-GAN~\cite{mogren2016c} is the first symbolic-domain music generative model that uses GAN~\cite{2014Generative}. In the generator and discriminator of this model, the recurrent network used is the long short-term memory(LSTM)~\cite{hochreiter1997long} network in order to model long-term dependency in symbolic music. HRNN~\cite{wu2019hierarchical} is a three-layer hierarchical RNN model, which models music from high-level to low-level by outputting bar profiles, beat profiles and notes in different hierarchical layers. MusicVAE~\cite{roberts2018hierarchical} is a model based on VAE~\cite{kingma2013auto}. This work introduced a novel sequential autoencoder and a hierarchical recurrent decoder to address the difficulty of modeling sequences with long-term dependency. It is not difficult to see that most of RNN-based symbolic music generative models design a kind of hierarchical structure to model the long-term and short-term structures in music.

\paragraph{CNN-based models}
The currently proposed CNN-based symbolic music generative models usually take advantage of GAN. These frameworks adopted convolutional neural network as the structure of discriminator and generator of GAN network. MidiNet~\cite{yang2017midinet} is the first GAN-based model with CNN as the core structure. MidiNet ingeniously adds two conditional inputs: 1-D condition and 2-D condition. In detail, 1-D condition is used to condition chords of the generation, while 2-D condition is the previous generated bar in order to condition the current bar. The introduction of these two conditions enhances the naturalness and continuity of the generated music. MuseGAN~\cite{dong2018musegan} is a GAN-based model for multi-track sequence generation. It proposed three GAN models: Jamming model, Composer model and Hybrid model. In addition, two models for modeling temporal relationships were also proposed. MuseGAN combines the aforementioned models and achieve high performance. Binary MuseGAN~\cite{dong2018convolutional} improved upon MuseGAN, which proposed to append to the generator an additional refiner, instead of using some post-processing approaches such as hard thresholding or Bernoulli sampling to achieve binary output. The proposed refiner has been proven to be beneficial to improve the performance of generated music. Compared to RNN-based models, the computational cost of CNN-based models is usually lower. As a result, CNN-based symbolic-domain music generative models can be trained more conveniently.

\tikzstyle{leaf}=[mybox,minimum height=1.2em,
fill=hidden-orange!50, text width=5em,  text=black,align=left,font=\footnotesize,
inner xsep=4pt,
inner ysep=1pt,
]

\begin{figure*}[thp]
 \centering

\begin{forest}
  forked edges,
  for tree={
  grow=east,
  reversed=true,  
  anchor=base west,
  parent anchor=east,
  child anchor=west,
  base=left,
  font=\normalsize,
  rectangle,
  draw=hiddendraw,
  rounded corners,
  align=left,
  minimum width=2.5em,
  inner xsep=4pt,
  inner ysep=0pt,
  },
  where level=1{text width=6em,font=\normalsize}{},
  where level=2{text width=6.9em,font=\normalsize}{},
  where level=3{font=\footnotesize,yshift=0.25pt}{},
    [Music generation
        [Symbolic
            [RNN-based
                [DeepBach~\cite{hadjeres2017deepbach}{,}MusicVAE~\cite{roberts2018hierarchical}\\
                C-RNN-GAN~\cite{mogren2016c}{,}HRNN~\cite{wu2019hierarchical}\\
                Song from PI~\cite{chu2016song}
                ,leaf,text width=12em
                ]
            ]
            [CNN-based
                [MidiNet~\cite{yang2017midinet}{,}MuseGAN~\cite{dong2018musegan}\\
                MuseGAN~\cite{dong2018convolutional}
                ,leaf,text width=12em
                ]
            ]
        ]
        [Audio
            [WaveNet~\cite{oord2016wavenet}{,}SampleRNN~\cite{mehri2016samplernn}\\
            WaveGAN~\cite{donahue2018adversarial}{,}MelGAN~\cite{kumar2019melgan}\\
            MelNet~\cite{vasquez2019melnet}{,}Jukebox~\cite{dhariwal2020jukebox}
            ,leaf,text width=12em
            ]
        ]
    ]
\end{forest}
\caption{A taxonomy of music generation.}
\label{main_taxonomy_of_music generation}
\end{figure*}
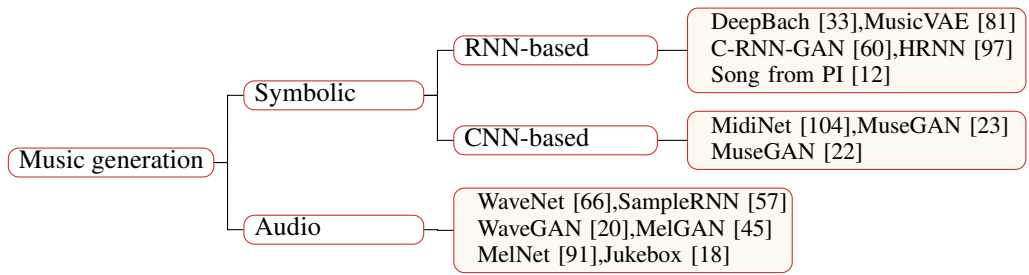

\subsection{Audio music generation}
\label{Audio music generation}
Different from symbolic music generation, approaches of audio music generation directly generate music at the raw audio sample point level, which allows for the generation of more diverse music and even vocals. So far, core structures of models for audio music generation include but are not limited to CNN and GAN. In the following, several neural network-based models for audio music generation will be introduced.

Note that several models used for TTS task are also capable of generating music in the raw audio domain, such as WaveNet, WaveGAN, MelGAN, MelNet and SampleRNN~\cite{oord2016wavenet,donahue2018adversarial,kumar2019melgan,vasquez2019melnet,mehri2016samplernn}. WaveNet~\cite{oord2016wavenet} is a versatile model that is competent for tasks such as multi-speaker speech generation, TTS, music generation and speech recognition. In terms of music generation, WaveNet adopts dilated casual convolutional layers to enlarge receptive field, which is crucial to obtain samples that sounded music. SampleRNN~\cite{mehri2016samplernn} can generate Beethoven-style piano sonatas at raw audio sample point level thanks to its hierarchy RNN structure. WaveGAN~\cite{donahue2018adversarial} improve the structure of DCGAN~\cite{radford2015unsupervised} to generate the sound of some instruments such as piano and drum. MelGAN~\cite{kumar2019melgan} is a GAN-based model that realizes music generation by decoding mel-spectrograms. MelNet~\cite{vasquez2019melnet} is also a generative model for music in the frequency domain, which is based on RNN.

Jukebox~\cite{dhariwal2020jukebox} is a model dedicated to audio music generation, which is based on VQ-VAE~\cite{razavi2019generating}. Jukebox is so powerful that it can even synthesize the singer’s vocals. To tackle the long context of raw audio, researchers use three separate VQ-VAE models with different timescales and make autoregressive Transformers as the key component of the encoders and decoders. Hence, Jukebox can generate high-fidelity and diverse songs in the raw audio domain with coherence up to multiple minutes. But the price is that it requires too much computational cost. Training a jukebox model requires hundreds of GPUs trained in parallel for several weeks. Similarly, reference also takes a lot of time and GPUs.

Although there is less research work on music generation in the raw audio domain, this area has attracted more and more attention from researchers. Audio music generative models not only does not require prior musical theory and a large number of human annotations, but also can generate more diverse music and even model singer’s vocal. However, audio noise and scratchiness that accompanies the music generated by these models is something we cannot ignore. In the future, we have to work more on the reduction of noise in audio music generative models and their computational cost.

\section{Audio-visual multimodal processing}
\label{Audio-visual multimodal processing}
Audio-visual multimodal processing means to accurately extract the low-dimensional features of acoustic and visual information, and then modelling the correlations between the two features to achieve some tasks such as lipreading. For computers, solving such problem is still a huge challenge due to the heterogeneous between audio and video. The audio and visual modalities are very different from each other in many aspects, for example, dimension and temporal resolution. In this review, we focus on four tasks related to audio-visual multimodal processing: lipreading, audio-visual speech separation, talking face generation and sound generation from video. They will be comprehensively introduced and summarized below.

\subsection{Lipreading}
\label{Lipreading}
Lipreading is an impressive skill to recognize what is being said from silent videos, and is a notoriously difficult task for humans to perform. Researchers take advantage of deep learning technology to train models to handle the lipreading task in recent years. Specifically, neural network-based lipreading models take silent facial(lip) videos as input, and output corresponding speech audio or characters. Several applications come to mind for automatic lipreading model: Enabling videoconferencing within a noisy environment; using surveillance video as a long-range listening device; facilitating conversation at a noisy party between people~\cite{ephrat2017vid2speech}. As a result, it is meaningful to develop automatic lipreading models, which bring convenience to our daily life. Several neural network-based lipreading models with high performance will be introduced below.

WLAS~\cite{chung2017lip} is a lipreading model that can transcribe videos of lip motion to characters, with or without the audio. CNN, LSTM~\cite{hochreiter1997long}, and Attention mechanism are adopted to model the correlations between the lip video and audio to generate the characters. The LRS dataset was also proposed by this work. LiRA~\cite{ma2021lira} is a self-supervised model, which projects the lip image sequence and audio waveform into the high-level representation respectively and the loss between them is calculated in the pretrain stage, then fine-tunes the model to achieve word-level and sentence-level lip reading. Ephrat et al.~\cite{ephrat2017improved} proposed that modeling the task as an acoustic regression problem has many advantages over visual-to-text modeling. For example, human emotions can be expressed in acoustic signal. Vid2Speech~\cite{ephrat2017vid2speech} is a CNN-based model that takes facial image sequences as input and outputs the corresponding speech raw audio waveform. ~\cite{ephrat2017improved} proposed a two-tower CNN-based model. One tower inputs facial grayscale images and the other inputs optical flow computed between facial image frames. Furthermore, the model adopts both mel-scale spectrograms and linear-scale spectrograms as the representation of audio waveform. ~\cite{assael2016lipnet,chung2017lip} also proposed models for lipreading task.

\subsection{Audio-visual speech separation}
\label{Audio-visual speech separation}
Humans are capable of focusing their auditory attention on a single sound source within a noisy environment, which is also known as the “cocktail party effect”. In recent years, the cocktail problem has become a typical problem in the field of computer speech recognition. Separating an input audio signal into its individual speech source is the definition of automatic speech separation. Ephrat et al.~\cite{ephrat2018looking} proposed that audio-visual speech separation can achieve higher performance than audio-only speech separation because viewing a speaker’s face enhances the model’s capacity to resolve perceived ambiguity. Automatic speech separation has many applications such as assistive technology for the hearing impaired and head-mounted assistive devices for noisy meeting scenarios. In this section, we focus on audio-visual speech separation and quantity of related work will be introduced below.

~\cite{ephrat2018looking} is the first attempt at speaker-independent audio-visual speech separation, which proposed a CNN-based model to encode the input facial images and speech audio spectrogram and output a complex mask for speech separation. In addition, the AVspeech dataset was also proposed in this work. AV-CVAE~\cite{nguyen2020deep} is an audio-visual speech separation model based on VAE~\cite{kingma2013auto}, which detects the lip movements of the speaker and predicts the individual separated speech audio. Acapella~\cite{montesinos2021cappella} is for audio-visual singing separation. The architecture is a two-stream CNN, which is called Y-Net for processing audio and video respectively. The Y-Net takes complex spectrogram and lip-region frames as input and outputs complex masks. In addition, a large dataset of solo singing videos was also proposed. VisualSpeech~\cite{gao2021visualvoice} takes a face image, an image sequence of lip movement and mixed audio as input, and outputs a complex mask. It also innovatively proposed the cross-modal embedding space to further correlate the information of audio and visual modalities. Facefilter~\cite{chung2020facefilter} uses still images as visual information and ~\cite{afouras2018conversation,gabbay2017visual} also proposed methods for audio-visual speech separation task.

\tikzstyle{leaf}=[mybox,minimum height=1.2em,
fill=hidden-orange!50, text width=5em,  text=black,align=left,font=\footnotesize,
inner xsep=4pt,
inner ysep=1pt,
]

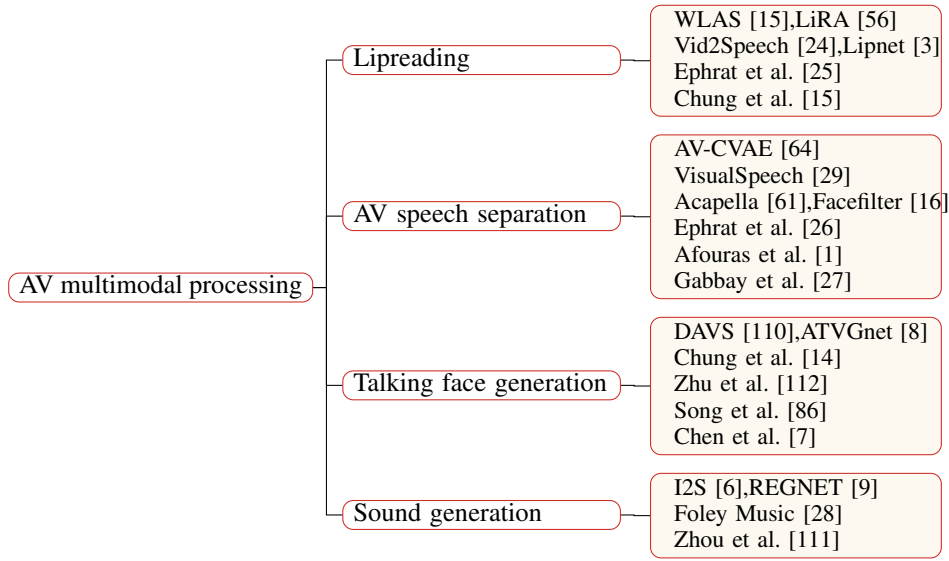
\begin{figure*}[thp]
 \centering

\begin{forest}
  forked edges,
  for tree={
  grow=east,
  reversed=true,  
  anchor=base west,
  parent anchor=east,
  child anchor=west,
  base=left,
  font=\normalsize,
  rectangle,
  draw=hiddendraw,
  rounded corners,
  align=left,
  minimum width=2.5em,
  inner xsep=4pt,
  inner ysep=0pt,
  },
  where level=1{text width=9.7em,font=\normalsize}{},
    [AV multimodal processing
        [Lipreading
            [WLAS~\cite{chung2017lip}{,}LiRA~\cite{ma2021lira}\\
            Vid2Speech~\cite{ephrat2017vid2speech}{,}Lipnet~\cite{assael2016lipnet}\\
            Ephrat et al{.}~\cite{ephrat2017improved}\\
            Chung et al{.}~\cite{chung2017lip}
            ,leaf,text width=10.2em
            ]
        ]
        [AV speech separation
            [AV-CVAE~\cite{nguyen2020deep}\\VisualSpeech~\cite{gao2021visualvoice}\\
            Acapella~\cite{montesinos2021cappella}{,}Facefilter~\cite{chung2020facefilter}\\
            Ephrat et al{.}~\cite{ephrat2018looking}\\
            Afouras et al{.}~\cite{afouras2018conversation}\\
            Gabbay et al{.}~\cite{gabbay2017visual}
            ,leaf,text width=10.2em
            ]
        ]
        [Talking face generation
            [DAVS~\cite{zhou2019talking}{,}ATVGnet~\cite{chen2019hierarchical}\\
            Chung et al{.}~\cite{chung2017you}\\
            Zhu et al{.}~\cite{zhu2018arbitrary}\\
            Song et al{.}~\cite{song2018talking}\\
            Chen et al{.}~\cite{chen2018lip}
            ,leaf,text width=10.2em
            ]
        ]
        [Sound generation
            [I2S~\cite{chen2017deep}{,}REGNET~\cite{chen2020generating}\\
            Foley Music~\cite{gan2020foley}\\
            Zhou et al{.}~\cite{zhou2018visual}
            ,leaf,text width=10.2em
            ]
        ]
    ]
\end{forest}
\caption{A taxonomy of audio-visual multimodal processing.}
\label{main_taxonomy_of_audio-visual multimodal processing}
\end{figure*}

\subsection{Talking face generation}
\label{Talking face generation}
Talking face generation means given a target face image and a clip of arbitrary speech audio, generating a natural talking face of the target character saying the given speech with lip synchronization. Meanwhile, the smooth transition of facial images should be guaranteed. It is a fresh and interesting task that has been an active topic. Generating talking face is challenging because the continuously changing facial region depends not only on visual information(given face image), but also on acoustic information(given speech audio), as well as achieving lip-speech synchronization. Meanwhile, it has many potential applications such as teleconferencing, generating virtual characters with specific facial movement and enhancing speech comprehension. A lot of work related to talking face generation task will be introduced below.

In previous efforts, researches performed 3D modeling of specific faces, then manipulating 3D meshes to generate the talking face. However, this method strongly relies on 3D modeling, which is time consuming, and cannot be extended to arbitrary identities. To our best knowledge, ~\cite{chung2017you} is the first work to use deep generative model for talking face generation. DAVS~\cite{zhou2019talking} is an end-to-end trainable deep neural network for talking face generation. The network firstly learns the joint audio-visual representation, then uses strategy of adversarial training to realize the latent space disentangling. ~\cite{chen2019hierarchical} proposed ATVGnet and the architecture of which can be divided into two parts: audio transformation network(AT-net) and visual generation network(VG-net) to process the acoustic and visual information respectively. In addition, a regression-based discriminator and a novel dynamically adjustable pixel-wise loss with an attention mechanism were also proposed. ~\cite{zhu2018arbitrary} proposed a novel talking face generative framework, which discovers audio-visual coherence via asymmetrical mutual information estimator. ~\cite{song2018talking,chen2018lip} also proposed methods for talking face generation, respectively.

\subsection{Sound generation from video}
\label{Sound generation from video}
The visual and auditory senses are arguably the two most important channels for humans to perceive the surrounding environment, and they are often closely associated with each other. From long-term observation of the natural environment, people can learn the association between audio and vision. In recent years, researchers have attempt leveraging learnable network to achieve this. Automatically generating sound from video has many application scenarios. For instance, combining videos with automatically generated sound to enhance the environment of virtual reality; adding background sound effects to videos automatically to avoid a lot of manual work. Several related studies will be introduced below.

~\cite{chen2017deep} is the first attempt to solve the audio-visual multimodal generation problem taking advantage of deep generative neural network. Two models named I2S and S2I based on CNN and GAN were proposed. These two models achieve sound-to-image and image-to-sound generation tasks, respectively. Moreover, the corresponding generation result of five representations of raw audio waveform were compared in the experiment part of this paper. In~\cite{zhou2018visual}, a SampleRNN~\cite{mehri2016samplernn}-style model was adopted as the sound generator, while three variants of the video encoder were proposed and compared. The VEGAS dataset containing 28109 cleaned videos spanning in 10 categories was also released. Foley Music~\cite{gan2020foley} is a Graph-Transformer based framework that learns to generate MIDI music from videos. The human pose features were extracted to capture the body movement, which was achieved by detecting the key points of body and hand. A Transformer-style decoder was also adopted. REGNET was proposed by~\cite{chen2020generating}, which introduced an information bottleneck to generate aligned sound from the given video.

In addition to the tasks mentioned above, leveraging the link between acoustic and visual information, multimodal learning methods explore a large range of interesting tasks such as face reconstruction from audio, emotion recognition, speaker identification. In the future, audio-visual multimodal processing will receive more attention from researchers within the area of deep learning.

\section{Datasets}
\label{Datasets}
In this section, several datasets that are widely adopted in TTS, music generation and audio-visual multimodal processing areas will be presented respectively. 

\subsection{Datasets for TTS}
\label{Datasets for TTS}

\paragraph{LJ speech}
The LJ speech dataset~\cite{ito2017lj} is a public speech dataset consists of 13100 short audio clips from a single speaker reading passages from 7 non-fiction books with a total length of approximately 24 hours. Clips vary in length from 1 to 10 seconds. The audio format is 16-bit PCM and the sample rate of audio is 22kHz.

\paragraph{VCTK}
The VCTK dataset~\cite{yamagishi2019cstr} includes speech data uttered by 110 English speakers with different accents. Each speaker reads about 400 sentences from a newspaper, the rainbow passage and an elicitation paragraph used for the speech accent archive. The audio format is 16-bit PCM with a sample rate of 44kHz.

\paragraph{LibriTTS corpus}
The LibriTTS dataset~\cite{zen2019libritts} is a multi-speaker English corpus, which consists approximately 585 hours of read English speech with a sample rate of 24kHz. The speech is split at sentence breaks and both normalize and original texts are included. Moreover, utterances with significant background noise are excluded and contextual information can be extracted.

\paragraph{Blizzard}
The Blizzard~\cite{prahallad2013blizzard} is a dataset for speech synthesis task, which consists of 315 hours English reading voice by a single female actor. The audio format is 16-bit PCM with a sample rate of 16kHz.

\subsection{Datasets for music generation}
\label{Datasets for music generation}

\paragraph{Lakh MIDI Dataset}
The Lakh MIDI Dataset~\cite{raffel2016learning} is a collection of 176581 unique midi-format files. 45129 files in this dataset have been matched and aligned to entries in the Million Song Dataset~\cite{bertin2011million}. The goal of the Lakh MIDI dataset is to facilitate large-scale music information retrieval, both symbolic and audio content-based.

\paragraph{Lakh Pianoroll Dataset}
The Lakh Pianoroll Dataset~\cite{dong2018musegan} consists of 174154 multitrack pianorolls, which are derived from the Lakh MIDI Dataset.

\paragraph{MAESTRO}
The MAESTRO~\cite{hawthorne2018enabling} is a dataset consists of 200 hours of paired audio and MIDI recordings from ten years of International Piano-e-Competition. Audio and MIDI files are aligned with about 3ms accuracy and sliced to individual pieces. Uncompressed audio can achieve CD-quality or higher and the audio format is 16-bit PCM with a sample rate of 44.1kHz-48kHz.

\tikzstyle{leaf}=[mybox,minimum height=1.2em,
fill=hidden-orange!50, text width=5em,  text=black,align=left,font=\footnotesize,
inner xsep=4pt,
inner ysep=1pt,
]

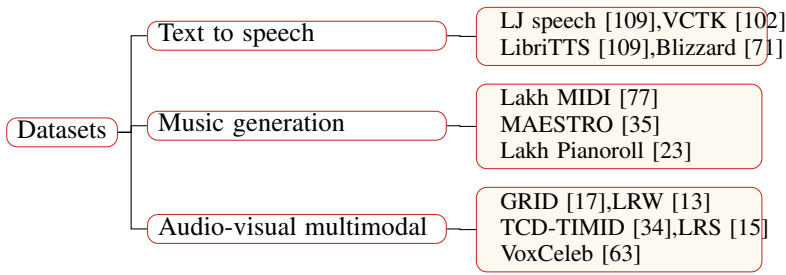
\begin{figure*}[thp]
 \centering

\begin{forest}
  forked edges,
  for tree={
  grow=east,
  reversed=true,  
  anchor=base west,
  parent anchor=east,
  child anchor=west,
  base=left,
  font=\normalsize,
  rectangle,
  draw=hiddendraw,
  rounded corners,
  align=left,
  minimum width=2.5em,
  inner xsep=4pt,
  inner ysep=0pt,
  },
  where level=1{text width=10.5em,font=\normalsize}{},
    [Datasets
        [Text to speech
            [LJ speech~\cite{zen2019libritts}{,}VCTK~\cite{yamagishi2019cstr}\\
            LibriTTS~\cite{zen2019libritts}{,}Blizzard~\cite{prahallad2013blizzard}
            ,leaf,text width=10.2em
            ]
        ]
        [Music generation
            [Lakh MIDI~\cite{raffel2016learning}\\MAESTRO~\cite{hawthorne2018enabling}\\
            Lakh Pianoroll~\cite{dong2018musegan}
            ,leaf,text width=10em
            ]
        ]
        [Audio-visual multimodal
            [GRID~\cite{cooke2006audio}{,}LRW~\cite{chung2016lip}\\
            TCD-TIMID~\cite{harte2015tcd}{,}LRS~\cite{chung2017lip}\\
            VoxCeleb~\cite{nagrani2017voxceleb}
            ,leaf,text width=10em
            ]
        ]
    ]
\end{forest}
\caption{A taxonomy of datasets.}
\label{main_taxonomy_of_datasets}
\end{figure*}

\subsection{Datasets for audio-visual multimodal processing}
\label{Datasets for audio-visual multimodal processing}

\paragraph{GRID}
The GRID audio-visual sentence corpus~\cite{cooke2006audio} is a large dataset of audio and facial video, which includes 1000 sentences spoken by 34 talkers. Talkers include 18 men and 16 women. Each sentence consists of a six-word sequence and the total vocabulary of this dataset is 51.

\paragraph{LRW}
The LRW dataset~\cite{chung2016lip} consists of about 1000 utterances containing 500 different words spoken by hundreds of different speakers. All videos in the dataset are 29 frames(1.16 second) in length and the word occurs in the middle of the video. The dataset is divided into training set, validation set and test set.

\paragraph{TCD-TIMID}
The TCD-TIMID dataset~\cite{harte2015tcd} is widely used in audio-video speech processing research, which consists of high-quality audio-video utterances from around 60 speakers. Every speaker uttering 98 sentences. 

\paragraph{LRS}
The LRS dataset~\cite{chung2017lip} is a dataset for visual speech recognition, which consists of over 100000 natural sentences from British television.

\paragraph{VoxCeleb}
The VoxCeleb dataset~\cite{nagrani2017voxceleb} is a large scale text-independent speaker identification dataset, which contains over 100000 utterances for 1251 celebrities. The videos are extracted from YouTube website. Speakers include various races, accents, ages and occupations, with a balanced gender distribution. The audio format is 16bit-PCM with a sample rate of 16 kHz.

\section{Conclusion}
\label{Conclusion}
In this review, we paid attention to the work on audio synthesis and audio-visual multimodal processing. The text-to-speech(TTS) and music generation tasks, which plays a crucial role in the maintenance of the audio synthesis field, were comprehensively summarized respectively. In the TTS task, two-stage and end-to-end methods were distinguished and introduced. As for the music generation task, symbolic domain and raw audio domain generative models were presented respectively. In the field of audio-visual multimodal processing, we focused on four typical tasks: lipreading, audio-visual speech separation, talking face generation and sound generation from video. The frameworks related to these tasks were introduced. Finally, several widely adopted datasets were also presented. Overall, this review provides considerable guidance to relevant researchers.

\bibliography{main}
\bibliographystyle{plainnat}

\end{document}